\documentclass[12pt]{article}

\usepackage{amsfonts}
\usepackage{epsfig,amssymb,euscript}
\usepackage{amsmath,amscd}

\numberwithin{equation}{section}
\begin{document}


\begin{titlepage}

\vfill

\vfill

\begin{center}

\baselineskip=16pt

{\Large\bf Enhanced Horizons}

\vskip 1.3cm

J.  B. Gutowski$^1$ and W. A. Sabra$^2$

\vskip 1cm

{\small{\it
$^1$DAMTP, Centre for Mathematical Sciences\\
University of Cambridge\\
Wilberforce Road, Cambridge, CB3 0WA, UK\\
Email: J.B.Gutowski@damtp.cam.ac.uk}}

\vskip .6cm {\small{\it
$^2$ Centre for Advanced Mathematical Sciences and Physics Department\\
American University of Beirut\\ Lebanon \\
Email: ws00@aub.edu.lb \\}}

\end{center}

\vfill

\begin{center}
\textbf{Abstract}
\end{center}
\begin{quote}

Half-supersymmetric geometries of $N=2$ five-dimensional gauged supergravity have
recently been fully classified using spinorial geometry techniques.
We use this classification to determine all possible regular
half-supersymmetric near-horizon geometries, assuming that
all of the gauge-invariant spinor bilinears are regular at the horizon. 
Four geometries are found, 
two of which have been found previously in \cite{minimalnear, harinear}.

\end{quote}
\vfill

\end{titlepage}


\section{Introduction}

Half supersymmetric solutions of gauged $N=2$, $D=5$ supergravity for which
at least one of the Killing spinors generates a timelike Killing vector,
were recently considered in \cite{halfgs, isometries}. The solutions found
fall into six classes. In all cases, the spacetime metrics are represented
by 
\begin{equation}
ds^{2}=f^{4}(dt+\Omega )^{2}-f^{-2}ds_{\mathbf{B}}^{2}
\end{equation}
where $ds_{\mathbf{B}}^{2}$ is the four dimensional base manifold. The
Killing vector $\partial /\partial t$ is a symmetry of the full solution, $f$
is a $t$-independent function and $\Omega $ is a $t$-independent 1-form on
the base manifold $\mathbf{B}$. In addition to the metric, the solutions are
also specified by scalar fields $X^{I}$ and Abelian gauge field strengths 
$F^{I}$. A summary of these solutions is given in the Appendix.

The starting point in the analysis of \cite{halfgs} is the construction of
Killing spinors as differential forms \cite{lawson, wang, harvey}. Gauge
symmetries are then employed in order to simplify the spinor as much as
possible; this approach was originally used to analyse higher dimensional
supergravity solutions in \cite{papadgran2005a, papadgran2005b,
papadgran2006a, papadgran2006b}. The conditions for the solutions to admit
one time-like Killing spinor 
\footnote{We refer to time-like Killing spinors as those spinors that generate
time-like Killing vectors as bilinears.} are then obtained. Such conditions
restrict the base manifold $\mathbf{B}$ to be K\"{a}hler. The half
supersymmetric solutions are then analysed by substituting the conditions
for the existence of one time-like Killing spinor into the generic Killing
spinor equations. We remark that all backgrounds preserving half of the
supersymmetry automatically solve all of the equations of motion, provided
the gauge fields satisfy the Bianchi identity. This is not the case for
solutions preserving only $1/4$ of the supersymmetry. Furthermore, the
Killing spinor equations, when expressed in terms of Dirac spinors, are
linear over $\mathbb{C}$. This implies that the allowed fractions of
preserved supersymmetries are $1/4$, $1/2,3/4$ or $1$. A great amount of
work has been devoted recently to classify and study all these solutions
(see \cite{electric}-\cite{harimulti}). The only maximally supersymmetric
solution is $AdS_{5}$ with vanishing gauge field strengths and constant
scalars.

In this paper we consider black hole solutions of $N=2,D=5$ gauged
supergravity coupled to an arbitrary number of Abelian vector multiplets. We
will assume that our black holes are supersymmetric and asymptotically 
$AdS_{5}$ with a single connected horizon. The reasoning given in 
\cite{reall:02} implies that such solutions have a time-like Killing spinor.
Therefore these solutions must at least preserve $1/4$ of the supersymmetry.
In addition, we assume that the solutions exhibit enhancement of
supersymmetry in the near horizon limit. As it has been shown that $3/4$
supersymmetric solutions are locally $AdS_{5}$ with vanishing gauge field
strengths, we shall therefore concentrate on half-supersymmetric solutions
classified in \cite{halfgs} and further simplified in \cite{isometries} and
we make extensive use of the results of these papers.

In our analysis, we shall assume that the scalar fields $X^{I}$ are smooth
in some neighbourhood of the horizon, and that all of the $Spin(4,1)$ and 
$U(1)$ -invariant bilinears constructed from the spinors are regular at the
horizon. The Killing vectors constructed from the Killing spinors can be
timelike or null.

If the bulk black hole geometry admits a Killing spinor, then the event horizon
should be preserved by the corresponding Killing vector, which must therefore be
tangential to the horizon at the horizon, and so has to become null at the horizon.
So the event horizon is a Killing horizon of this Killing vector. However,
not all of the spinors associated with the near-horizon geometry can be extended to
give spinors in the bulk geometry away from the horizon, though we shall assume that
at least two of the four spinors in the near horizon geometry generate a Killing vector
which corresponds to the timelike Killing vector in the bulk, for which the event horizon is
a Killing horizon. In particular, this means that in general, the Killing vectors associated
with the additional spinors in the near-horizon  geometry do not have to preserve the horizon,
as these spinors do not generate isometries of the bulk geometry. However, in the analysis
presented here, we shall assume that all of the Killing vectors obtained as Killing spinor
bilinears become null at the horizon.
Using these constraints, we obtain necessary conditions for the half
supersymmetric solutions of \cite{halfgs} to describe regular near horizon
geometries.  It is found that the constraints are incompatible with five of
the six  classes of solution found in \cite{halfgs}, and hence these classes
cannot describe near horizon geometries. The remaining class of  solution,
described in section 6
provides four possible near horizon geometries. Two of these solutions, given in
({\ref{oldsol1}}) and ({\ref{oldsol2}}) have already
been found in \cite{minimalnear, harinear}. The other two solutions
are given in ({\ref{newmet1}}) and ({\ref{newmet2}}).

Our work is organised as follows. In section 2, we study the regularity of
gauge invariant spinor bilinears. This immediately excludes two classes of
solutions of \cite{halfgs} as horizon geometries. In sections 3, 4, 5 we
investigate the constraints that the existence of a horizon imposes on three
other classes of solutions in \cite{halfgs}, and prove that these solutions
cannot contain regular horizons. In section 6, the remaining class of
solutions is analysed, and four types of possible near horizon geometry are
constructed. In section 7, we analyse some aspects of the causal structure
of the two solutions found in section 6 which were not obtained in the
classification in \cite{minimalnear, harinear}. In section 8 we present our conclusions.

\section{Spinorial Regularity}

We follow the notation of \cite{halfgs} and denote the Killing spinors by
complexified forms on $\mathbb{R}^{2}$. In particular, 
\begin{eqnarray}
\label{kilsp1}
\epsilon ^{1} &=&f,\quad \epsilon ^{2}=fe^{12},\quad   \nonumber \\
\eta ^{1} &=&\lambda +\mu ^{p}e^{p}+\sigma e^{12},\quad \eta ^{2}=-\sigma
^{\ast }-\epsilon _{ij}(\mu ^{i})^{\ast }e^{j}+\lambda ^{\ast }e^{12}.
\end{eqnarray}
The horizon $\mathbf{H}$ corresponds to the hypersurface $f=0$. Let 
$\mathcal{B}$ be the $Spin(4,1)$ invariant inner product on spinors given in 
\cite{preons}. We shall impose a \textquotedblleft spinorial
regularity\textquotedblright\ condition on the solutions. The spinor
components transform under gauge transformations and thus cannot
individually be taken to be regular at the horizon. We shall require that
all gauge invariant spinor bilinears are regular at the horizon.

In particular, this implies that $\mathcal{B}(\epsilon ^{1},\epsilon ^{2})$, 
$\mathcal{B}(\eta ^{1},\eta ^{2})$, $\mathcal{B}(\epsilon ^{1},\eta ^{2})-
\mathcal{B}(\epsilon ^{2},\eta ^{1})$, $\mathcal{B}(\epsilon ^{1},\eta ^{1})+
\mathcal{B}(\epsilon ^{2},\eta ^{2})$ are regular. Moreover, observe that
these scalars are \textit{both} $Spin(4,1)$ invariant due to the $Spin(4,1)$
invariance of $\mathcal{B}$, and also $U(1)$ invariant; where we recall that
under $U(1)$ transformations, symplectic Majorana Killing spinors $\eta
^{1},\eta ^{2}$ transform as 
\begin{eqnarray}
\eta ^{1} &\rightarrow &\cos \theta \eta ^{1}-\sin \theta \eta ^{2},  \nonumber
\\
\eta ^{2} &\rightarrow &\sin \theta \eta ^{1}+\cos \theta \eta ^{2}
\end{eqnarray}
where $\theta \in \mathbb{R}$, so that $\eta ^{1}+i\eta ^{2}\rightarrow
e^{i\theta }(\eta ^{1}+i\eta ^{2})$.

Observe also that $\mathcal{B}(\epsilon ^{1}+i\epsilon ^{2},\eta ^{1}+i\eta
^{2})\mathcal{B}(\epsilon ^{1}-i\epsilon ^{2},\eta ^{1}-i\eta ^{2})$ is 
$Spin(4,1)\times U(1)$ invariant. Evaluating these scalars in the basis used
in \cite{halfgs}, we observe that $f^2$, $|\lambda|^2+|\sigma|^2
-|\mu^1|^2-|\mu^2|^2$, $f(\lambda+\lambda^*)$, $f(\sigma+\sigma^*)$ and 
$f^2((\lambda-\lambda^*)^2+(\sigma-\sigma^*)^2)$ must be regular at the
horizon. Hence $f^{2}(|\lambda |^{2}+|\sigma |^{2})$ and 
$f^{2}(|\mu^{1}|^{2}+|\mu ^{2}|^{2})$ must also be regular at the horizon.

Furthermore, recall that the spinors $\epsilon ^{1},\epsilon ^{2},\eta
^{1},\eta ^{2}$ generate two globally well-defined Killing vectors with
components $\mathcal{B}(\epsilon ^{1},\gamma ^{\mu }\epsilon ^{2})$ and 
$\mathcal{B}(\eta ^{1},\gamma ^{\mu }\eta ^{2})$. These Killing vectors have
norms $f^{4}$ and $(|\lambda |^{2}+|\sigma |^{2}-|\mu ^{1}|^{2}-|\mu
^{2}|^{2})^{2}$ respectively; so they are therefore either timelike or null.
We assume that the Killing vector spinor bilinear generated by $\epsilon^1, \epsilon^2$ 
extends to give an isometry of the bulk geometry, which must preserve the horizon.
Hence
\begin{equation}
f \rightarrow 0
\end{equation}
at the horizon. We shall further assume that the Killing vector associated with 
$\eta^1, \eta^2$ also becomes null at the horizon, so
\begin{eqnarray}
|\lambda |^{2}+|\sigma |^{2}-|\mu ^{1}|^{2}-|\mu ^{2}|^{2} \rightarrow 0,
\end{eqnarray}
at the horizon as well. We remark that both these limits must hold, as a consequence of the reasoning set out in
\cite{reall:02}, if one assumes that not only the near-horizon geometry, but also the black hole bulk geometry preserves half the supersymmetry.

 Next, recall from \cite{halfgs}
that on defining the real vector field $K$ on the K\"ahler base ${\bf{B}}$ by
\begin{equation}
K^p = i f^2 \mu^p, \qquad K^{\bar{p}} = -i f^2 (\mu^p)^*
\end{equation}
one has 
\begin{equation}
K^{2}= 2 K_p K^p = 2f^{4}(|\mu ^{1}|^{2}+|\mu ^{2}|^{2})=
2f^{2}.f^{2}(|\mu ^{1}|^{2}+|\mu^{2}|^{2}).  \label{reg1}
\end{equation}
Hence $K^{2}\rightarrow 0$ as $f\rightarrow 0$.

These conditions can be used to immediately exclude solutions of type $(4)$,
because they have $f=1$, and also solutions of type $(6)$ are excluded,
because for these solutions it has been shown that $K^{2}$ is a non-zero
constant (see the Appendix). Hence it remains to examine solutions of type 
$(1)$, $(2)$, $(3)$ and $(5)$.

\section{Near-Horizon Analysis of Type $(1)$ Solutions}

For these solutions, we first recall some useful relations presented in 
\cite{halfgs}. The $v$ co-ordinate is related to $K^{2}$ via \cite{halfgs} 
\begin{equation}
v={\frac{K^{2}}{\sqrt{2}c}}  \label{vcoord}
\end{equation}
and hence, at the horizon, $v\rightarrow 0$. Furthermore, $\cos Y$ and 
$\sin Y$ are defined in terms of the spinor components via 
\begin{equation}
\cos Y={\frac{\mathrm{Re\ }(\sigma ^{2}+(\lambda ^{\ast })^{2})}{|\lambda
|^{2}+|\sigma |^{2}}},\qquad \sin Y={\frac{\mathrm{Im\ }(\sigma
^{2}+(\lambda ^{\ast })^{2})}{|\lambda |^{2}+|\sigma |^{2}}}
\end{equation}
with $\lambda ,\sigma $ constrained via 
\begin{equation}
\mathrm{Im\ }(\lambda \sigma )=0.
\end{equation}
To proceed, it is convenient to define $\Xi $ by 
\begin{equation}
H^{2}=c^{2}v^{2}f^{-6}+\Xi  \label{xieq}
\end{equation}
where $H^2$ is defined in \cite{halfgs} as
\begin{equation}
H^2 = f^{-2} K^2 (|\lambda|^2+|\sigma|^2) \ .
\end{equation}

Note also that 
\begin{equation}
{\frac{\partial \Xi }{\partial v}}=-\theta \cos Y,\qquad {\frac{\partial \Xi 
}{\partial u}}=-\theta H^{2}v\sin ^{2}Y.  \label{xder}
\end{equation}
These conditions are obtained by using ({\ref{Hcons2}})
to compute $dH$, and then using ({\ref{bc1}}) and  ({\ref{Hcons2}})
to compute $df$, using the conditions on $X_I$ implied by ({\ref{bc1}})
together with the identity $X^I dX_I=0$ which follows from the constraints of
Very Special geometry.

Next observe that 
\begin{equation}
\Xi =2f^{2}(|\mu ^{1}|^{2}+|\mu ^{2}|^{2})(|\lambda |^{2}+|\sigma |^{2}-|\mu
^{1}|^{2}-|\mu ^{2}|^{2})
\end{equation}
and hence $\Xi $ is regular, and $\Xi \rightarrow 0$ at the horizon. There
are then two possible cases, according as to whether the constant $\theta $
is zero or nonzero.

\subsection{Solutions with $\protect\theta \neq 0$}

\bigskip In the first case $\theta \neq 0$. Then ({\ref{bc1}}) together with ({\ref{xieq}})
can be used to rewrite the conditions on the scalars as

\begin{equation}
f^{-2}X_{I}=-{\frac{\chi }{c}}\left( {\frac{\Xi }{\theta v}}+{1}\right)
V_{I}+{\frac{q_{I}}{v}}
\end{equation}
for constant $q_{I}$. This expression implies, on contracting with $X^I$ and using the condition
$X_I X^I=1$ obtained from the Very Special geometry of the scalar manifold, that
\begin{equation}
f^{-2}(1+{\frac{\sqrt{2}\chi }{\theta }}V_{I}X^{I}(|\lambda |^{2}+|\sigma
|^{2}-|\mu ^{1}|^{2}-|\mu ^{2}|^{2}))=\left( {\frac{q_{I}}{v}}- {\frac{\chi 
}{c}}V_{I}\right) X^{I} \ .
\end{equation}
This implies that at the horizon 
\begin{equation}
vf^{-2}\rightarrow h
\end{equation}
where $h$ is a regular function.

Furthermore, one also finds that 
\begin{equation}
vf^{-2}\left( X_{I}+{\frac{\sqrt{2}\chi }{\theta }}(|\lambda |^{2}+|\sigma
|^{2}-|\mu ^{1}|^{2}-|\mu ^{2}|^{2})V_{I}\right) =-{\frac{\chi }{c}}
vV_{I}+q_{I}
\end{equation}
so that if $X_{Ihor}$ denotes the restriction of $X_{I}$ to the horizon, 
\begin{equation}
hX_{Ihor}=q_{I}.
\end{equation}
If $h=0$ at any point of the horizon then $q_{I}=0$ for all $I$. In this
case 
\begin{equation}
X_{I}=-{\frac{\chi }{c}}f^{2}(1+{\frac{\Xi }{\theta v}})V_{I}.
\end{equation}
Then from the constraints of the Very Special geometry, 
\begin{equation}
f^{2}\left( \theta ^{-1}{\frac{\Xi }{v}}+1\right) =\delta 
\end{equation}
for constant $\delta $. However, note that $f^{2}\Xi v^{-1}\rightarrow 0$ at
the horizon, and so $\delta =0$. This implies that 
\begin{equation}
\Xi =-\theta v.
\end{equation}
But then ${\frac{\partial \Xi }{\partial u}}=0$, which implies that $\sin Y=0
$; however for this class of solutions, $\sin Y \neq 0$ (see the Appendix, also
the original derivation of the solutions in \cite{halfgs}) . Hence $h\neq 0$ on
the horizon, and we can write 
\begin{equation}
X_{Ihor}=h^{-1}q_{I}
\end{equation}
on the horizon. The constraints of very special geometry force $h$ to be
constant at the horizon, and hence without loss of generality, we can set 
$h=1$ on the horizon, so that 
\begin{equation}
X_{Ihor}=q_{I}.
\end{equation}
Next consider $vH^{2}$, note that one can write 
\begin{equation}
vH^{2}={\frac{2\sqrt{2}}{c}}f^{2}(|\lambda |^{2}+|\sigma |^{2})(f^{2}(|\mu
^{1}|^{2}+|\mu ^{2}|^{2}))^{2}
\end{equation}
hence $vH^{2}$ is regular at the horizon. Furthermore, 
\begin{equation}
vH^{2}=\Xi v+c^{2}v^{3}f^{-6}\rightarrow c^{2}
\end{equation}
at the horizon. In addition, 
\begin{equation}
\cos Y-1={\frac{c}{\sqrt{2}}}\left( {\frac{f^{2}(\sigma -\sigma ^{\ast
})^{2}+f^{2}(\lambda -\lambda ^{\ast })^{2}}{vH^{2}.f^{4}v^{-2}}}\right) 
\end{equation}
so it follows that $\cos Y$ is also globally well-defined and regular in
some neighbourhood of the horizon.

To proceed, recall that we have shown that the scalars $X_{I}$ should be
constant on the horizon. However, we are assuming that the
half-supersymmetric solution we are investigating is \textit{already} the
near-horizon limit of some black hole (or ring) solution, so taking the near
horizon limit of the scalars twice will not alter them, i.e. we must take 
$X_{I}=q_{I}$ for our solution. This implies that 
\begin{equation}
-{\frac{\chi }{c}}f^{2}({\frac{\Xi }{\theta v}}
+1)V_{I}+(f^{2}v^{-1}-1)q_{I}=0.
\end{equation}

If $f^{2}v^{-1}=0$ everywhere then $\Xi =-\theta v$ and again 
${\frac{\partial \Xi }{\partial u}}=0$ implies that $\sin Y=0$, in contradiction
with our assumption that $\sin Y\neq 0$. Hence, there is a (non-zero) real
constant $\delta $ such that $q_{I}=\delta V_{I}$, and so 
\begin{equation}
-{\frac{\chi }{c}}f^{2}({\frac{\Xi }{\theta v}}+1)+\delta (f^{2}v^{-1}-1)=0.
\end{equation}
Note that $X^{I}V_{I}=\delta ^{-1}$.

It is convenient to rewrite ({\ref{xder}}) using the expression 
\begin{equation}
\Xi = - \theta v + {\frac{\delta c \theta }{\chi}} (1-f^{-2} v)
\end{equation}

to give 
\begin{eqnarray}
{\frac{\partial F}{\partial v}} &=&{\frac{\chi }{\delta c}}(\cos Y-1), 
\nonumber \\
{\frac{\partial F}{\partial u}} &=&{\frac{\chi }{\delta c}}\sin
^{2}Y(c^{2}F^{3}-\theta v^{2}+{\frac{\delta c\theta }{\chi }}v(1-F)),
\label{Fder}
\end{eqnarray}
and 
\begin{eqnarray}
{\frac{\partial Y}{\partial v}} &=&-{\frac{\sin Y}{(c^{2}F^{3}-\theta v^{2} +
{\frac{\delta c\theta }{\chi }}v(1-F))}}(3\chi c\delta ^{-1}F^{2}-\theta v),
\nonumber \\
{\frac{\partial Y}{\partial u}} &=&\sin Y(2\chi c\delta ^{-1}F^{2}+\theta v+
{\frac{\delta c\theta }{\chi }}(F-1))
\nonumber \\
&+&\sin Y\cos Y(3\chi 
c\delta^{-1}F^{2}-\theta v),  \nonumber \\
&&  \label{Yder}
\end{eqnarray}
where we have set $F=vf^{-2}$. $F$ is regular in the near-horizon geometry
and $F\rightarrow 1$ at the horizon.

The equations ({\ref{Fder}}) and ({\ref{Yder}}) imply that 
\begin{eqnarray}
{\frac{\partial F }{\partial v^2}}(c^2 F^3 - \theta v^2 + {\frac{\delta c
\theta }{\chi}}v(1-F)) = (-2{\frac{\partial F }{\partial v}}-{\frac{\delta c 
}{\chi}}\big({\frac{\partial F }{\partial v}}\big)^2) (3 \chi c \delta^{-1}
F^2 - \theta v)
\nonumber \\
\end{eqnarray}
which can be integrated to give 
\begin{eqnarray}
{\frac{\partial F }{\partial v}}(c^2 F^3 - \theta v^2 + {\frac{\delta c
\theta }{\chi}}v(1-F)) -{\frac{\delta c \theta }{\chi}} (F-{\frac{1 }{2}}
F^2)+2 \chi c \delta^{-1} F^3 = G(u)
\nonumber \\
\end{eqnarray}
for some function $G(u)$. Acting on this expression with 
${\frac{\partial }{\partial u}}$ we find that $G$ must be constant. Set $G=k$.

We therefore find 
\begin{equation}
{\frac{\partial F}{\partial v}}={\frac{(k+{\frac{\delta c\theta }{\chi }}(F-
{\frac{1}{2}}F^{2})-2\chi c\delta ^{-1}F^{3})}{(c^{2}F^{3}-\theta v^{2}+
{\frac{\delta c\theta }{\chi }}v(1-F))}}  \label{Fspecial1}
\end{equation}
and note further that ({\ref{Fder}}) also implies that 
\begin{eqnarray}
{\frac{\partial F}{\partial u}} &=&{\frac{(4c\chi ^{2}F^{3}-2\delta
^{2}c\theta (F-{\frac{1}{2}}F^{2})-2k\chi \delta )}{4\chi ^{2}\delta
(-c^{2}\chi F^{3}+\theta \chi v^{2}-\delta c\theta v(1-F))}}  \nonumber \\
&\times &(-2\delta ^{2}c^{2}\theta (F-{\frac{1}{2}}F^{2})+4\chi \delta
c\theta v(F-1)+4\theta \chi ^{2}v^{2}-2ck\chi \delta ).  \label{Fspecial2}
\end{eqnarray}

One can use ({\ref{Fspecial2}}) to change co-ordinates from $(u,v)$ to 
$(F,v) $ and hence obtain an explicit expression for the metric. Setting 
$F=1+y$ we find

\bigskip

\begin{eqnarray}
ds^{2} &=&{\frac{v^{2}}{(1+y)^{2}}}dt^{\prime }{}^{2}-{\frac{1}{\chi
c(1+y)^{2}}}\bigg(2c^{2}\chi (1+y)^{3}-\theta \chi v^{2}-\delta c\theta
vy\bigg)dt^{\prime }d\phi   \nonumber \\
&&+{\frac{1}{\chi ^{2}(1+y)^{2}}}\bigg( (4c\chi ^{2}v+2\delta c^{2}\chi
y)(1+y)^{3}-(\delta ^{2}c\theta +2k\chi \delta )v\bigg)dt^{\prime }dw  \nonumber \\
&&-{\frac{\delta \bigg( 2k\chi +\delta c\theta \bigg) \bigg(4c\chi ^{2}(1+y)^{3}-\delta
^{2}c\theta -2k\chi \delta \bigg)}{4\chi ^{4}(1+y)^{2}}}dw^{2}  \nonumber \\
&&+{\frac{\theta }{2c\chi ^{3}(1+y)^{2}}}\bigg(-(\delta ^{2}c\theta +2k\delta
\chi )(\delta cy+\chi v)
\nonumber \\
&& \qquad \qquad \qquad +(2\chi ^{2}c^{2}\delta y+4\chi
^{3}cv)(1+y)^{3}\bigg)d\phi dw  \nonumber \\
&&+{\frac{(1+y)}{c^{2}\delta ^{2}\theta (y^{2}-1)-2ck\chi \delta +4\chi
\theta v(\delta cy+\chi v)}} \bigg( 4\chi ^{2}dv^{2}+4\delta c\chi dvdy \bigg) \nonumber \\
&&+{\frac{4\chi c\delta ^{2}(1+y)\bigg( c^{2}\chi (1+y)^{3}-\theta \chi
v^{2}-\delta c\theta vy \bigg)}{(c^{2}\delta ^{2}\theta (y^{2}-1)-2ck\chi \delta
+4\chi \theta v(\delta cy+\chi v))}}  \nonumber \\
&&\ \  \times {\frac{1}{(4c\chi ^{2}(1+y)^{3}-2k\chi \delta +\delta ^{2}c\theta
(y^{2}-1))}}dy^{2}
\nonumber \\
&&+{\frac{\theta ^{2}(\chi v+\delta cy)^{2}}{4\chi ^{2}c^{2}(1+y)^{2}}}d\phi
^{2} \ .
\end{eqnarray}

Next consider the Killing vector ${\frac{\partial }{\partial t^{\prime }}}- 
{\frac{1}{2c}}{\frac{\partial }{\partial \phi }}$. It is straightforward to
show that the norm of this vector tends to $1$ as $v\rightarrow 0$ and 
$y\rightarrow 1$, so the Killing vector is timelike at the horizon; this then
implies that this geometry cannot contain an event horizon.

\bigskip

\subsubsection{Solutions with $\protect\theta=0$}

For solutions with $\theta =0$, $\Xi $ is constant. However, the requirement
that $\Xi \rightarrow 0$ at the horizon fixes $\Xi =0$ everywhere, and hence $H^2=c^2 v^2 f^{-6}$. It is
then straightforward to integrate up equations ({\ref{Hcons2}}) and ({\ref{Ycons2}}) (setting $\theta=0$), to find 
\begin{equation}
v^{-3}f^{6}=\delta (1+\cos Y)
\end{equation}
for constant $\delta >0$. Next note that 
\begin{equation}
\cos Y+1={\frac{1}{\sqrt{2}c}}v^{-1}f^{2}(f^{2}(\lambda +
\lambda ^{\ast})^{2}+f^{2}(\sigma +\sigma ^{\ast })^{2})  \label{coseq1}
\end{equation}
which implies that 
\begin{equation}
\sqrt{2}c\delta ^{-1}v^{-2}f^{4}=f^{2}(\lambda +\lambda ^{\ast
})^{2}+f^{2}(\sigma +\sigma ^{\ast })^{2}
\end{equation}
and hence $v^{-2}f^{4}$ is regular (and in particular bounded above) in some
neighbourhood of the horizon.

Next recall that 
\begin{equation}
v f^{-2} = {\frac{\sqrt{2} }{c}} f^2 (|\mu^1|^2+|\mu^2|^2)
\end{equation}
which is also regular at the horizon. In order for $v^{-2} f^4$ to be
regular, $v f^{-2}$ cannot vanish at any point of the horizon. Hence $v^{-1}
f^2$ is also regular in some neighbourhood of the horizon. Therefore, from 
({\ref{coseq1}}), $\cos Y$ is also regular in some neighbourhood of the
horizon.

It is also straightforward to show that from the constraints given in the Appendix that
\begin{equation}
\left( {\frac{\partial }{\partial v}}+{\frac{\delta }{c^{2}}}{\frac{\partial 
}{\partial u}}\right) (vf^{-2}X_{I})=0.
\end{equation}
Defining 
\begin{equation}
y=v-{\frac{c^{2}}{\delta }}u,\quad z=v+{\frac{c^{2}}{\delta }}u,
\end{equation}
we observe that $vf^{-2}$ and $X_{I}$ depend only on $y$ and not $z$. The
remaining content of ({\ref{scalcon1}}) can be then written as 
\begin{equation}
{\frac{d}{dy}}(FX_{I})=-{\frac{\chi }{c}}V_{I}(2-\delta ^{-1}F^{-3})
\end{equation}
where $F=vf^{-2}$. Observe that this equation implies that 
\begin{equation}
X_{I}=\left( {\frac{X_{\tilde{I}}}{V_{\tilde{I}}}}\right) V_{I}+
{\frac{q_{I}}{V_{\tilde{I}}F}.}  \label{xfix2}
\end{equation}
This equation, together with the constraints of the Very Special geometry,
can be used to fix the function ${\frac{X_{\tilde{I}}}{V_{\tilde{I}}}}$ in
terms of $F$, and hence $X_{I}$ in terms of $F$. It remains to compute the
spacetime metric: we find 
\begin{eqnarray}
ds^{2} &=&{\frac{v^{2}}{F^{2}}}dt^{\prime }{}^{2}-2cFdt^{\prime }(d\phi +
\big[-{\frac{cF}{\chi }}\big({\frac{X_{\tilde{I}}}{V_{\tilde{I}}}}\big)
+v(-2+\delta ^{-1}F^{-3})\big]dw)  \nonumber \\
&&-{\frac{c^{2}}{\delta }}F(2-\delta ^{-1}F^{-3})dw^{2}-
{\frac{2\delta F}{c^{2}}}dv^{2}-{\frac{2\delta F}{c\chi V_{I}X^{I}}}dvdF  \nonumber \\
&&-{\frac{\delta F}{(\chi V_{I}X^{I})^{2}(2-\delta ^{-1}F^{-3})}}dF^{2}.
\end{eqnarray}

Now consider the Killing vector ${\frac{\partial }{\partial t}}  
-{\frac{1 }{2}} {\frac{\partial }{\partial \phi}}$. This has norm $v^2 F^{-2} +cF$ which
tends to $cF$ on the horizon. However, by the definition of the co-ordinate 
$v$ in terms of $K^2$ in ({\ref{vcoord}}), it follows that $cF$ is positive
everywhere on the horizon, so this Killing vector is timelike on the
horizon. Again, this implies that this geometry cannot contain an event
horizon.

\section{Near Horizon Analysis of Type $(2)$ Solutions}

For these solutions, the co-ordinate $v$ is again related to $K^{2}$ via 
\begin{equation}
v={\frac{K^{2}}{\sqrt{2}c}}
\end{equation}
so $v\rightarrow 0$ at the horizon. The scalars satisfy 
\begin{equation}
X_{I}=f^{2}\left( -{\frac{2\chi }{c}}V_{I}+{\frac{\rho _{I}}{K^{2}}}\right)
\end{equation}
for constants $\rho _{I}$. Suppose that $\rho _{I}=0$ for all $I$. Then 
$X_{I}=-{\frac{2\chi }{c}}f^{2}V_{I}$. The constraints of Very Special
geometry imply that $f^{2}$ must be constant. However, $f^{2}\rightarrow 0$
at the horizon, so $f=0$ everywhere, in contradiction to our initial
assumption. Hence, as not all $\rho _{I}$ are vanishing, and the scalars are
regular in some neighbourhood of the horizon, it follows that 
${\frac{f^{2}}{K^{2}}}$ must also be regular in some neighbourhood of the horizon.
Furthermore, 
\begin{equation}
{\frac{K^{2}}{f^{2}}}=2f^{2}(|\mu ^{1}|^{2}+|\mu ^{2}|^{2})
\end{equation}
which is also regular. It follows that both ${\frac{f^{2}}{K^{2}}}$ and 
${\frac{K^{2}}{f^{2}}}$ are regular, and non-vanishing in some neighbourhood
of the horizon.

Moreover, we find that $X_{I}\rightarrow {\frac{f^{2}}{K^{2}}}\rho _{I}$ at
the horizon. This implies that ${\frac{f^{2}}{K^{2}}}$ is constant on the
horizon, without loss of generality, we can set ${\frac{f^{2}}{K^{2}}}=1$ on
the horizon, so that $X_{I}=\rho _{I}$ on the horizon. However, by
assumption, $X_{I}$ is \textit{already} in its near-horizon limit, so we set 
$X_{I}=\rho _{I}$. Hence we find 
\begin{equation}
-{\frac{2\chi }{c}}V_{I}=\rho _{I}\left( {\frac{1}{f^{2}}}-{\frac{1}{K^{2}}}
\right) .
\end{equation}
As not all $\rho _{I}$ are zero, there exists a constant $\delta $ such that 
\begin{equation}
{\frac{1}{f^{2}}}-{\frac{1}{K^{2}}}=\delta
\end{equation}
so that $-{\frac{2\chi }{c}}V_{I}=\delta \rho _{I}$. If $\delta =0$, then 
$V_{I}=0$ for all $I$, in contradiction to the assumption that not all $V_{I}$
vanish. Hence $\delta \neq 0$. Note that 
\begin{equation}
K^{2}={\frac{f^{2}}{1-\delta f^{2}}}
\end{equation}
and hence 
\begin{equation}
f^{2}={\frac{\sqrt{2}cv}{1+\sqrt{2}\delta cv}.}
\end{equation}
We then obtain the spacetime metric explicitly; if $\theta \neq 0$ then 
\begin{eqnarray}
ds^{2} &=&{\frac{2c^{2}v^{2}}{(1+\sqrt{2}\delta cv)^{2}}}dt^{2}- 
{\frac{\sqrt{2}}{\theta }}(1+\sqrt{2}\delta cv)dt\sigma ^{1}
-{\frac{1}{\sqrt{2}c\theta }} (1+\sqrt{2}\delta cv)(\sigma ^{1})^{2}  \nonumber \\
&-&{\frac{1}{\sqrt{2}|c\theta |}}(1+\sqrt{2}\delta cv)((\sigma
^{2})^{2}+(\sigma ^{3})^{2})
\nonumber \\
&-&\left( {\frac{1+\sqrt{2}\delta cv}{\sqrt{2}
c(\theta v^{2}+{\frac{1}{2\sqrt{2}c}}(1+\sqrt{2}\delta cv)^{3})}}\right)
dv^{2},  \nonumber \\
&&
\end{eqnarray}
and if $\theta =0$, 
\begin{eqnarray}
ds^{2} &=&{\frac{2c^{2}v^{2}}{(1+\sqrt{2}\delta cv)^{2}}}dt^{2}+2\sqrt{2}
c^{3}(1+\sqrt{2}\delta cv)dt\sigma ^{1}  \nonumber \\
&-&\sqrt{2}c^{2}(1+\sqrt{2}\delta cv)\left( (\sigma ^{2})^{2}+(\sigma
^{3})^{2}\right) -{\frac{2}{(1+\sqrt{2}\delta cv)^{2}}}dv^{2}.
\end{eqnarray}
In all cases, it is straightforward to see that one can choose an
appropriate linear combination of ${\frac{\partial }{\partial t}}$ and  
$\sigma ^{1}$ in order to obtain a Killing vector which has a positive,
non-vanishing norm in the limit $v\rightarrow 0$. Hence these geometries
cannot contain an event horizon.

\section{Near Horizon Analysis of Type $(3)$ Solutions}

For these solutions, the co-ordinate $v$ is again related to $K^2$ via

\begin{equation}
v = {\frac{K^2 }{\sqrt{2} c}}
\end{equation}
so $v\rightarrow 0$ at the horizon. Also recall that the scalars $X^{I}$ are
constant. Then the constraint

\begin{equation}
-{\frac{\theta }{K^{2}}}-{\frac{\sqrt{2}c}{K^{2}}}f^{-2}(|\lambda
|^{2}+|\sigma |^{2})+3\sqrt{2}\chi f^{-4}V_{I}X^{I}+{\frac{c}{\sqrt{2}}}
f^{-6}=0
\end{equation}
can be rewritten (taking without loss of generality $K^{2}=f^{2}$) as

\begin{equation}
-\theta K^{2}-\sqrt{2}c(|\lambda |^{2}+|\sigma |^{2}-|\mu ^{1}|^{2}-|\mu
^{2}|^{2})+3\sqrt{2}\chi V_{I}X^{I}=0.
\end{equation}
Taking the limit of the LHS as one approaches the horizon, we obtain the
constraint

\begin{equation}
V_{I}X^{I}=0
\end{equation}
hence we note that these solutions cannot arise in the minimal gauged
supergravity. It is convenient to define

\begin{equation}
\Lambda =c\theta -{\frac{9}{\sqrt{2}}}\chi ^{2}Q^{IJ}V_{I}V_{J}.
\end{equation}
Then if $\Lambda \neq 0$, the metric is

\begin{eqnarray}
ds^{2} &=&2c^{2}v^{2}dt^{2}+{\frac{\sqrt{2}c}{\Lambda }}dt\sigma ^{1}+ 
{\frac{c\theta }{\sqrt{2}\Lambda ^{2}}}(\sigma ^{1})^{2}  \nonumber \\
&-&{\frac{1}{{\frac{1}{2}}-\sqrt{2}c\theta v^{2}}}dv^{2}-{\frac{1}{\sqrt{2}
|\Lambda |}}((\sigma ^{2})^{2}+(\sigma ^{3})^{2})
\end{eqnarray}
and if $\Lambda =0$,

\begin{equation}
ds^{2}=2c^{2}v^{2}dt^{2}+2cdt\sigma ^{1}+\sqrt{2}c\theta (\sigma ^{1})^{2}- 
{\frac{1}{{\frac{1}{2}}-\sqrt{2}c\theta v^{2}}}dv^{2}-((\sigma
^{2})^{2}+(\sigma ^{3})^{2}).
\end{equation}
Again, in all cases, one can choose an appropriate linear combination of 
${\frac{\partial }{\partial t}}$ and $\sigma ^{1}$ in order to obtain a
Killing vector which has a positive, non-vanishing norm in the limit 
$v\rightarrow 0$. Hence these geometries cannot contain an event horizon.

\section{Near Horizon Analysis of Type $(5)$ Solutions}

For these solutions, 
\begin{equation}
K^{2}={\frac{1}{\varrho ^{2}}}e^{\sqrt{2}\varrho ^{2}\psi }
\end{equation}
and hence $\psi \rightarrow -\infty $ at the horizon. To proceed, we must
consider the cases for which $(\mathrm{Im}\lambda )^{2}+(\mathrm{Im\ }\sigma
)^{2}\neq 0$, or $\mathrm{Im\ }\lambda =\mathrm{Im\ }\sigma =0$ separately.

\subsection{Solutions with $( \mathrm{Im \ } \protect\lambda)^2+( \mathrm{Im
\ } \protect\sigma)^2 \neq 0$}

To proceed with the analysis of solutions for which $(\mathrm{Im\ }\lambda
)^{2}+(\mathrm{Im\ }\sigma )^{2}\neq 0$, note that there exist constants 
$c_{3},c_{4}$ such that $c_{3}^{2}+c_{4}^{2}=\varrho ^{2}\neq 0$, and one
obtains constraints on the spinor components \cite{halfgs}: 
\begin{equation}
c_{3}(\lambda +\lambda ^{\ast })+c_{4}(\sigma +\sigma ^{\ast })=2\varrho
^{2}tue^{{\frac{1}{\sqrt{2}}}\varrho ^{2}\psi },
\end{equation}
and 
\begin{equation}
c_{4}(\lambda +\lambda ^{\ast })-c_{3}(\sigma +\sigma ^{\ast })=
{\frac{uQ}{\varrho ^{2}}}e^{-{\frac{1}{\sqrt{2}}}\varrho ^{2}\psi }.
\end{equation}
It follows that 
\begin{eqnarray}
|\mu ^{1}|^{2}+|\mu ^{2}|^{2}-|\lambda |^{2}-|\sigma |^{2} &=&
{\frac{1}{4}}e^{-\sqrt{2}\varrho ^{2}\psi }\left[ {\frac{2u^{-4}}{\varrho ^{2}}}
+u^{2}(\mathcal{G}^{2}+\mathcal{H}^{2})-{\frac{u^{2}Q^{2}}{\varrho ^{6}}}\right]  
\nonumber \\
&&-\varrho ^{2}t^{2}u^{2}e^{\sqrt{2}\varrho ^{2}\psi }.  \label{secondinv}
\end{eqnarray}
Therefore 
\begin{equation}
|\mu ^{1}|^{2}+|\mu ^{2}|^{2}-|\lambda |^{2}-|\sigma |^{2}=
-{\frac{\xi }{4\varrho ^{6}}}u^{2}e^{-\sqrt{2}\varrho ^{2}\psi }-
\varrho ^{2}t^{2}u^{2}e^{\sqrt{2}\varrho ^{2}\psi }.
\end{equation}
Next note that 
\begin{equation}
K^{2}f^{-2}=2f^{2}(|\mu ^{1}|^{2}+|\mu ^{2}|^{2})=
{\frac{u^{-2}}{\varrho ^{2}}.}
\end{equation}
Hence $u^{-2}$ must be regular at the horizon.

Consider the constraints on the scalars: 
\begin{equation}
u^{-2}X_{I}={\frac{\chi }{\varrho ^{4}}}QV_{I}+q_{I},
\end{equation}
as the left hand side of this expression is regular at the horizon, and not
all of the $V_{I}$ vanish, it follows that $Q$ must also be regular at the
horizon. Then from ({\ref{q2con1}}) we see that $\mathcal{G}^{2}+
\mathcal{H}^{2}$ must also be regular at the horizon. Furthermore, as 
\begin{equation}
c_{4}f(\lambda +\lambda ^{\ast })-c_{3}f(\sigma +\sigma ^{\ast })=
{\frac{1}{\varrho ^{2}}}u^{2}Q,
\end{equation}
$u^{2}Q$ must also be regular.

To proceed further, note that 
\begin{equation}
X_{I}=u^{2}\left( {\frac{\chi }{\varrho ^{4}}}QV_{I}+q_{I}\right) .
\end{equation}
Then there are two possibilities. If at least one $q_{I}\neq 0$, then
regularity of the scalars $X_{I}$ and of $u^{2}Q$ implies that $u^{2}$ is
also regular at the horizon. In the second possibility, $q_{I}=0$ for all $I$. 
Then 
\begin{equation}
X_{I}={\frac{\chi }{\varrho ^{4}}}u^{2}QV_{I}
\end{equation}
which together with the constraints of Very Special geometry, implies that 
$u^{2}Q$ is a (non-zero) constant, and the scalars $X_{I}$ are also constant,
with $V_{I}X^{I}\neq 0$, and 
\begin{equation}
1={\frac{\chi }{\varrho ^{4}}}u^{2}QV_{I}X^{I}.
\end{equation}
In order to obtain additional constraints on the function $u$, observe that 
\begin{eqnarray}
Q_{IJ}F_{\mu \nu }^{I}F^{J\mu \nu } &=&2u^{2}\varrho ^{2}+9\chi
^{2}Q^{IJ}V_{I}V_{J}\left( -{\frac{1}{\varrho ^{4}}}Q^{2}u^{6}+4+ 
{\frac{1}{\varrho ^{4}}}\xi u^{6}\right)  \nonumber \\
&&-12\chi V_{I}X^{I}Qu^{4}+2u^{8}Q^{2}+\xi u^{8}.  \label{f2con2b}
\end{eqnarray}
This expression must be regular at the horizon, as a consequence of the
Einstein equations (together with the assumption that the scalars are
regular at the horizon).

We have already shown that $u^{-2}$ is regular near the horizon. Suppose
then that we are in the case for which $Q u^2$ and the scalars $X_I$ are
constant. Suppose further that $u^{-2} \rightarrow 0$ at some point of the
horizon. Then there must exist a sequence of points $p_n$ tending towards
the horizon such that $u(p_n) \rightarrow \infty$ as $n \rightarrow \infty$.
So consider ({\ref{f2con2b}}) at these points. The LHS must be regular at
the horizon, however the RHS diverges; if $\xi \neq 0$ then the divergence
is as $u^8$, whereas if $\xi=0$ then the divergence is as $u^4$. In both
cases there is a contradiction. Hence $u^{-2}$ is bounded below by a nonzero
positive number on the horizon. It follows that $u^2$ is also regular near
the horizon.

Finally, note that if $\xi>0$ then the RHS of (\ref{secondinv}) tends to 
$-\infty$ at the horizon, whereas the LHS tends to zero. Hence we must have 
$\xi \leq 0$.

Having obtained these results, we are now in a position to write down the
near-horizon metrics in Gaussian Null co-ordinates; let $\xi = -L^2$. For
the case with $L > 0$, make the following co-ordinate transformations:

\begin{eqnarray}
r &=&{\frac{e^{\sqrt{2}\varrho ^{2}\psi }}{Lu^{2}},}  \nonumber \\
t^{\prime } &=&t+{\frac{1}{2\varrho ^{4}u^{2}r},}  \nonumber \\
z^{\prime } &=&z-{\frac{\varrho }{\sqrt{2}L}}\log (u^{2}r).
\end{eqnarray}
In the new co-ordinates, the metric is

\begin{eqnarray}
\label{oldsol1}
ds^{2} &=&L^{2}u^{8}r^{2}(dt^{\prime })^{2}+2dt^{\prime }dr  \nonumber \\
&+&2rdt^{\prime }\left[ 2u^{-1}{\frac{du}{dQ}}dQ-{\frac{Lu^{6}}{\sqrt{2}%
\varrho ^{2}}}(Qd\phi -\varrho ^{-3}(2\varrho ^{4}u^{-6}-L^{2})dz^{\prime })%
\right]   \nonumber \\
&-&{\frac{u^{4}L^{2}}{2\varrho ^{10}}}(-L^{2}+2\varrho
^{4}u^{-6})(dz^{\prime }-{\frac{Q\varrho ^{3}}{(-L^{2}+2\varrho ^{4}u^{-6})}}%
d\phi )^{2}  \nonumber \\
&+&{\frac{u^{-2}}{(-L^{2}+2\varrho ^{4}u^{-6})}}(Q^{2}+L^{2}-2\varrho
^{4}u^{-6})d\phi ^{2}
\nonumber \\
&+&{\frac{1}{2}}u^{-2}\varrho ^{-4}(Q^{2}+L^{2}-2\varrho
^{4}u^{-6})^{-1}dQ^{2} \ . \nonumber \\
&&
\end{eqnarray}
Although it appears that $\varrho $ is a free parameter of the solution, one
can without loss of generality set $\varrho =1$. To see this, make the
rescalings 
\begin{eqnarray}
u &=&\varrho ^{-2}{\hat{u}},\qquad q_{I}=\varrho ^{4}{\hat{q}}_{I},\qquad
Q=\varrho ^{8}{\hat{Q},}  \nonumber \\
L &=&\varrho ^{8}{\hat{L}},\qquad \phi =\varrho ^{-2}\hat{\phi},\qquad
z^{\prime }=\varrho ^{-7}{\hat{z}}^{\prime },
\end{eqnarray}
and then drop the ${\hat{}}\ $, one then obtains the solution with $\varrho
=1$. In the case for which the scalar manifold is symmetric, this solution
has been been found in \cite{harinear}, it is the \textquotedblleft
non-static" near horizon geometry with non-constant scalars. To see this,
recall that when the scalar manifold is symmetric, one has the identity 
\begin{equation}
{\frac{9}{2}}C^{IJK}X_{I}X_{J}X_{K}=1
\end{equation}
where 
\begin{equation}
C^{IJK}=\delta ^{II^{\prime }}\delta ^{JJ^{\prime }}\delta ^{KK^{\prime
}}C_{I^{\prime }J^{\prime }K^{\prime }}\ .
\end{equation}

It is then possible to construct the metrics explicitly. As mentioned
previously, we shall set $\varrho =1$ without loss of generality. To
proceed, it is convenient to set 
\begin{equation}
\xi ^{3}={\frac{9}{2}}C^{IJK}V_{I}V_{J}V_{K}
\end{equation}
and we assume that $\xi \neq 0$. Also define $K_{I}$, $x$ by 
\begin{equation}
q_{I}={\frac{2}{C^{2}}}K_{I},\quad Q={\frac{2}{\xi \chi C^{2}}}x,
\end{equation}
where $C>0$ is constant. Next, define 
\begin{eqnarray}
{\hat{\alpha}}_{0} &=&{\frac{9}{2}}C^{IJN}K_{I}K_{J}K_{N},  \nonumber \\
{\hat{\alpha}}_{1} &=&{\frac{9}{2\xi }}C^{IJN}K_{I}K_{J}V_{N},  \nonumber \\
{\hat{\alpha}}_{2} &=&{\frac{9}{2\xi ^{2}}}C^{IJN}K_{I}V_{J}V_{N}\ ,
\end{eqnarray}
so that 
\begin{equation}
u^{-6}={\frac{1}{8}}C^{6}H
\end{equation}
where 
\begin{equation}
H=x^{3}+3{\hat{\alpha}}_{2}x^{2}+3{\hat{\alpha}}_{1}x+{\hat{\alpha}}_{0}\ .
\end{equation}
Then, on defining 
\begin{equation}
L={\frac{4\Delta _{0}}{C^{4}}},\qquad z^{\prime }=-{\frac{1}{4\sqrt{2}\Delta
_{0}}}C^{6}x^{1},\qquad \phi =-\sqrt{2}\xi \chi x^{2},\qquad t^{\prime }=-v,
\end{equation}
where $\Delta _{0}>0$ is constant, we recover the metric found in 
\cite{harinear} 
{\footnote{Up to a constant shift in the $x$ co-ordinate, which can be used to set 
${\hat{\alpha}}_{2}=0$}}.

In the special case when $L=0$, it is convenient to define $R$ by 
\begin{equation}
t=e^{-\sqrt{2}\varrho ^{2}\psi }R
\end{equation}
and note that ({\ref{secondinv}}) implies that 
\begin{equation}
f^{2}(|\mu ^{1}|^{2}+|\mu ^{2}|-|\lambda |^{2}-|\sigma |^{2})=-\varrho
^{2}R^{2}
\end{equation}
so it follows that $R\rightarrow 0$ at the horizon. The metric in the new
co-ordinates is given by 
\begin{eqnarray}
ds^{2} &=&u^{4}(dR-\sqrt{2}\varrho ^{2}Rd\psi )^{2}-{\frac{\sqrt{2}}{\varrho
^{2}}}u^{4}(dR-\sqrt{2}\varrho ^{2}Rd\psi )(Qd\phi -2\varrho
u^{-6}dz)\nonumber \\ &-&u^{-2}d\psi ^{2}  
+({\frac{1}{2}}\varrho ^{-4}u^{4}Q^{2}-u^{-2})d\phi ^{2}-{\frac{1}{2}}%
\varrho ^{-8}u^{-2}(2u^{-6}-\varrho ^{-4}Q^{2})^{-1}dQ^{2}.  \nonumber \\ \label{newmet1}
\end{eqnarray}
Again, one can rescale and set $\varrho =1$ without loss of generality; the
rescalings are given by

\begin{eqnarray}
u &=&\varrho ^{-2}{\hat{u}},\qquad q_{I}=\varrho ^{4}{\hat{q}}_{I},\qquad
Q=\varrho ^{8}{\hat{Q}},\qquad \phi =\varrho ^{-2}\hat{\phi},  \nonumber \\
\psi &=&\varrho ^{-2}\hat{\psi},\qquad R=\varrho ^{4}{\hat{R}},\qquad
z=\varrho ^{-7}\hat{z},
\end{eqnarray}
and on dropping the ${\hat{}}\ $, one obtains the solution with $\varrho =1$.

\subsection{Solutions with $( \mathrm{Im \ } \protect\lambda)^2=( \mathrm{Im
\ } \protect\sigma)^2 =0$}

When $\mathrm{Im\ }\lambda =\mathrm{Im\ }\sigma =0$, and set 
$r=e^{\sqrt{2}\varrho ^{2}\psi }$; one obtains

\begin{eqnarray}
ds^{2} &=&r^{2}\left[ dt-{\frac{3\chi V_{I}X^{I}}{\sqrt{2}\varrho ^{2}r}}
(d\phi +\beta )\right] ^{2}-(d\phi +\beta )^{2}-
{\frac{1}{2\varrho ^{4}r^{2}}}dr^{2}-ds^{2}(M)  \nonumber \\
&&  \label{imvanishmet1}
\end{eqnarray}
where the scalars $X^{I}$ are constant, and the constraints on $ds^{2}(M)$
and $\beta $ are given in the Appendix. It is convenient to define 
\begin{equation}
\delta ={\frac{3\chi }{\sqrt{2}\varrho ^{2}}}V_{I}X^{I}.
\end{equation}

Note that in order for the metric ({\ref{imvanishmet1}}) to describe an
event horizon, ${\frac{\partial }{\partial \phi }}$ cannot be timelike at
the horizon; this implies that $\delta ^{2}-1\leq 0$. Consider first the
case $\delta ^{2}<1$. On defining the co-ordinates $t^{\prime },\phi
^{\prime }$ by 
\begin{eqnarray}
t^{\prime } &=&t+{\frac{\sqrt{1-\delta ^{2}}}{\sqrt{2}\varrho ^{2}r},} 
\nonumber \\
\phi ^{\prime } &=&\phi +{\frac{\delta }{\sqrt{2}\varrho ^{2}\sqrt{1-\delta
^{2}}}}\log r,
\end{eqnarray}
one can write the metric in Gaussian null co-ordinates: 
\begin{eqnarray}
\label{oldsol2}
ds^{2} &=&r^{2}(dt^{\prime })^{2}-2\delta rdt^{\prime }(d\phi ^{\prime
}+\beta )+{\frac{\sqrt{2}}{\varrho ^{2}\sqrt{1-\delta ^{2}}}}dt^{\prime }dr 
\nonumber \\
&+&(\delta ^{2}-1)(d\phi ^{\prime }+\beta )^{2}-ds^{2}(M).
\end{eqnarray}

This class of solutions corresponds to the second class of \textquotedblleft
non-static" solutions found in \cite{harinear}, when the scalars are
constant. It is straightforward to see that the geometries are isometric, by
making the identification 
\begin{equation}
t^{\prime }=-\sqrt{2}\varrho ^{2}\sqrt{1-\delta ^{2}}v
\end{equation}
and setting 
\begin{equation}
\Delta =\sqrt{2}\varrho ^{2}\sqrt{1-\delta ^{2}},\qquad Z^{1}=\sqrt{1-\delta
^{2}}(d\phi ^{\prime }+\beta ),\qquad \beta ={\frac{1}{\sqrt{1-\delta ^{2}}}}
\alpha ,
\end{equation}
and noting that 
\begin{equation}
\Delta ^{2}+g^{2}\lambda =2\varrho ^{4}+9\chi
^{2}(Q^{IJ}-2X^{I}X^{J})V_{I}V_{J}={\frac{1}{2}}{}^{(M)}R\ .
\end{equation}

In the special case of $\delta ^{2}=1$, it is convenient to define $R$ by 
\begin{equation}
t=\delta e^{-\sqrt{2}\varrho ^{2}\psi }R
\end{equation}
and define $\phi ^{\prime }$ by 
\begin{equation}
\phi ^{\prime }=\phi -{\frac{R}{2}.}
\end{equation}

It is then straightforward to show that 
\begin{equation}
f^{2}(|\mu ^{1}|^{2}+|\mu ^{2}|-|\lambda |^{2}-|\sigma |^{2})=-\varrho
^{2}R^{2}
\end{equation}
so $R\rightarrow 0$ at the horizon, and the metric in the new co-ordinates
is 
\begin{equation}
ds^{2}=(dR-\sqrt{2}\varrho ^{2}Rd\psi )(-2(d\phi ^{\prime }+\beta )-
 \sqrt{2}\varrho ^{2}Rd\psi )-d\psi ^{2}-ds^{2}(M).  \label{newmet2}
\end{equation}

\section{Causal Structure}

In this section, we analyse the causal structure of the two classes of
possible near horizon metric ({\ref{newmet1}}) and ({\ref{newmet2}}) which
do not correspond to solutions constructed in \cite{minimalnear, harinear}.

To proceed, consider causal geodesics in the spacetime given by 
({\ref{newmet2}}) which pass through the horizon. As 
${\frac{\partial }{\partial \phi ^{\prime }}}$ is a Killing vector, one finds that 
\begin{equation}
\dot{R}-\sqrt{2}\varrho ^{2}R\dot{\psi}=k_{1}  \label{geod1}
\end{equation}
for constant $k_{1}$, where $\dot{}={\frac{d}{d\tau }}$, and $\tau $ is the
affine parameter. We assume that the geodesic passes through the event
horizon in a finite affine parameter. Then the $R$ component of the geodesic
equation can be written as 
\begin{eqnarray}
{\frac{d}{d\tau }}\big(-2(\dot{\phi}^{\prime }+\hat{\beta})-\sqrt{2}\varrho
^{2}R\dot{\psi}\big)+\sqrt{2}\varrho
^{2}k_{1}\dot{\psi} \nonumber \\ \qquad  \qquad +\sqrt{2}\varrho ^{2}\dot{\psi}\big(-2( 
\dot{\phi}^{\prime }+\hat{\beta})-\sqrt{2}\varrho ^{2}R\dot{\psi}\big)=0  \nonumber \\
\end{eqnarray}
where $\hat{\beta}$ denotes the restriction of the 1-form $\beta $ to the
worldline of the geodesic. Hence, 
\begin{equation}
-2(\dot{\phi}^{\prime }+\hat{\beta})-\sqrt{2}\varrho ^{2}R\dot{\psi}
=-k_{1}+k_{2}e^{-\sqrt{2}\varrho ^{2}\psi }  \label{geod2}
\end{equation}
for constant $k_{2}$. Next, the $\psi $ component of the geodesic equation
implies 
\begin{equation}
4\varrho ^{4}R^{2}\dot{\psi}+2\sqrt{2}\varrho ^{2}R(\hat{\phi}^{\prime }+ 
\hat{\beta})-2\dot{\psi}-\sqrt{2}\varrho ^{2}R\dot{R}=k_{3}
\end{equation}
for constant $k_{3}$. Using ({\ref{geod2}}) this constraint can be
simplified to 
\begin{equation}
-2\dot{\psi}-\sqrt{2}\varrho ^{2}k_{2}Re^{-\sqrt{2}\varrho ^{2}\psi }=k_{3} \ .
\label{geod3}
\end{equation}
It is straightforward to see that $k_{3}\neq 0$, because $k_{3}=0$ implies
that $-2\dot{\psi}=k_{3}$. However, this is not possible, because one must
have $\psi \rightarrow -\infty $ at the horizon. Hence, on combining 
({\ref{geod3}}) and ({\ref{geod1}}), we obtain 
\begin{equation}
{\ddot{\psi}}=-{\frac{1}{\sqrt{2}}}\varrho ^{2}k_{1}k_{2}e^{-\sqrt{2}\varrho
^{2}\psi }.  \label{geocon1}
\end{equation}
If $V$ is the tangent vector to the geodesic, then 
\begin{equation}
V^{2}=-k_{1}^{2}-\dot{\psi}^{2}+k_{1}k_{2}e^{-\sqrt{2}\varrho ^{2}\psi
}-|V_{M}^{2}|,  \label{geocon2}
\end{equation}
where $V_{M}$ denotes the portion of the tangent vector pointing in the
directions corresponding to the 2-manifold $M$. Hence, for causal geodesics,
one must have $k_{1}k_{2}>0$ (otherwise, one is forced to take $\dot{\psi}=0$, 
in contradiction to the fact that $\psi \rightarrow -\infty $ at the
horizon). On integrating ({\ref{geocon1}}) we find 
\begin{equation}
-\dot{\psi}^{2}+k_{1}k_{2}e^{-\sqrt{2}\varrho ^{2}\psi }=k_{4},
\label{geocon3}
\end{equation}
for constant $k_{4}$. Moreover, as 
\begin{equation}
V^{2}=-k_{1}^{2}+k_{4}-|V_{M}^{2}|,
\end{equation}
it follows that for causal geodesics, one must have $k_{4}>0$ (otherwise,
one must have $k_{1}=0$, again giving a contradiction). Then ({\ref{geod3}})
implies that 
\begin{equation}
R={\frac{1}{\sqrt{2}\varrho ^{2}k_{2}}}e^{\sqrt{2}\varrho ^{2}\psi } \big(
-k_{3}\pm 2\sqrt{k_{1}k_{2}e^{-\sqrt{2}\varrho ^{2}\psi }-k_{4}}\big)
\end{equation}
so, as expected, $R\rightarrow 0$ as $\psi \rightarrow -\infty $. Finally,
on integrating ({\ref{geocon3}}) one finds that 
\begin{equation}
\psi ={\frac{\sqrt{2}}{\varrho ^{2}}}\log \bigg(\sqrt{\frac{k_{1}k_{2}}{k_{4}}} 
\sin \big(\sqrt{\frac{\varrho ^{4}k_{4}}{2}}(\tau -\tau _{0})\big)\bigg)
\end{equation}
for constant $\tau _{0}$ and hence 
\begin{eqnarray}
R &=&{\frac{1}{\sqrt{2}\varrho ^{2}}}{\frac{k_{1}}{k_{2}}}\bigg(-k_{3}\sin
^{2}\big(\sqrt{\frac{\varrho ^{4}k_{4}}{2}}(\tau -\tau _{0})\big)  \nonumber \\
&\pm &2\sqrt{k_{4}}\sin \big(\sqrt{\frac{\varrho ^{4}k_{4}}{2}}(\tau -\tau
_{0})\big)\cos \big(\sqrt{\frac{\varrho ^{4}k_{4}}{2}}(\tau -\tau _{0})\big) 
\bigg).
\end{eqnarray}
The geodesic passes through the horizon when $\sin \big(\sqrt{\frac{\varrho
^{4}k_{4}}{2}}(\tau -\tau _{0})\big)=0$.

Next, consider the geodesics of the metric ({\ref{newmet1}}). We set 
$\varrho=1$ without loss of generality. As ${\frac{\partial }{\partial \phi}}$, 
${\frac{\partial }{\partial \psi}}$ and ${\frac{\partial }{\partial z}}$
are Killing vectors, there are three constants $k_1, k_2, k_3$ satisfying

\begin{equation}
u^{-2}(\dot{R}-\sqrt{2}R\dot{\psi})=k_{1},  \label{geob1}
\end{equation}
\begin{equation}
-\sqrt{2}Qk_{1}u^{6}+(u^{4}Q^{2}-2u^{-2})\dot{\phi}=k_{2},  \label{geob2}
\end{equation}
\begin{equation}
-2\sqrt{2}Ru^{6}k_{1}+2Ru^{4}(Q\dot{\phi}-2u^{-6}\dot{z})-2u^{-2}\dot{\psi}
=k_{3},  \label{geob3}
\end{equation}
and the $R$ component of the geodesic equation can be integrated up to give 
\begin{equation}
-{\frac{1}{\sqrt{2}R}}(k_{3}+2u^{-2}\dot{\psi})=k_{4}e^{-\sqrt{2}\psi },
\label{geob4}
\end{equation}
for constant $k_{4}$. On evaluating the norm $V^{2}$ of the tangent vector
to the geodesic one finds 
\begin{eqnarray}
V^{2} &=&-k_{1}^{2}u^{8}+k_{1}k_{4}u^{2}e^{-\sqrt{2}\psi }-u^{-2} \dot{\psi}^{2}  
\nonumber \\
&+&{\frac{1}{2}}{\frac{(k_{2}+\sqrt{2}k_{1}Qu^{6})^{2}}{u^{4}Q^{2}-2u^{-2}}}
+ {\frac{1}{2}}u^{-2}{\frac{\dot{Q}^{2}}{Q^{2}-2u^{-6}}.}  \label{norm2}
\end{eqnarray}
In addition, as the Killing vector ${\frac{\partial }{\partial \phi }}$ must
be either null or spacelike on the horizon, $Q^{2}-2u^{-6}$ cannot be
positive at any point of the horizon. Hence, for the geodesic to be causal,
we must have $k_{1}k_{4}>0$ ($k_{1}k_{4}=0$ would force $\dot{\psi}=0$,
which is not possible, as $\psi \rightarrow -\infty $ at the horizon).

It is then straightforward to integrate up ({\ref{geob1}}) and ({\ref{geob4}}) 
to find 
\begin{equation}
R={\frac{e^{\sqrt{2}\psi }}{\sqrt{2}k_{4}}}\bigg(-k_{3}\pm \sqrt{4k_{1}k_{4}e^{-\sqrt{2}\psi }
+k_{3}^{2}+2\sqrt{2}k_{4}k_{5}}\bigg),
\end{equation}
for constant $k_{5}$. Moreover, we also find 
\begin{equation}
k_{1}k_{4}u^{2}e^{-\sqrt{2}\psi }-u^{-2}\dot{\psi}^{2}=-{\frac{1}{4}}
u^{2}(k_{3}^{2}+2\sqrt{2}k_{4}k_{5}).  \label{geob5}
\end{equation}
Hence, by comparing with ({\ref{norm2}}), it is clear that for causal
geodesics, one must have $k_{3}^{2}+2\sqrt{2}k_{4}k_{5}<0$. Finally, it is
useful to make a change of parameter from $\tau $ to $y$, where 
\begin{equation}
{\frac{dy}{d\tau }}=u^{2}\ ,
\end{equation}
from the properties of $u$ derived in the previous section, this change of
parameter is well-defined in some neighbourhood of the horizon. Then 
({\ref{geob5}}) can be rewritten as 
\begin{equation}
\left( {\frac{d\psi }{dy}}\right) ^{2}=k_{1}k_{4}e^{-\sqrt{2}\psi } +{\frac{1}{4}}(k_{3}^{2}
+2\sqrt{2}k_{4}k_{5}),
\end{equation}
which can be integrated up. One finds that the dependence of $\psi $ and $R$
on $y$ is (up to redefinition of constants), the same as the $\tau $-dependence 
of the metric ({\ref{newmet2}}). Finally, to determine the
behaviour of $Q$ along the geodesic, note that 
\begin{eqnarray}
\left( {\frac{dQ}{dy}}\right) ^{2} &=&\big(2V^{2}u^{-2}+{\frac{1}{2}}
(k_{3}^{2}+2\sqrt{2}k_{4}k_{5})\big)(Q^{2}-2u^{-6})  \nonumber \\
&-&(k_{2}^{2}u^{-6}+2\sqrt{2}k_{1}k_{2}Q+4k_{1}^{2}).
\end{eqnarray}

Suppose that the scalar manifold is symmetric. It is straightforward to show
that $u$ cannot vanish along any timelike geodesic. This is because, if $u
\rightarrow 0$, then $Q \sim u^{-2}$, and the RHS of the above expression
diverges as $-4 V^2 u^{-8}$, whereas the LHS is non-negative. The same holds
for those null geodesics for which $k_3^2 +2 \sqrt{2} k_4 k_5 + k_2^2>0$.

\section{Conclusions}

We have derived all possible half-supersymmetric regular near horizon black
hole solutions in $N=2$ five-dimensional gauged supergravity,
subject to the assumption that, in the near-horizon limit, the event horizon is a
Killing horizon of both Killing vectors generated from the Killing spinors
$(\epsilon^1, \epsilon^2)$ and $(\eta^1, \eta^2)$, and that all of the Killing spinor
bilinears are regular at the horizon. There are four
geometries. Two of the geometries have already been found in 
\cite{minimalnear, harinear}. The other two geometries are given by:

\begin{itemize}
\item[i)] 
\begin{equation}
ds^{2}=(dR-\sqrt{2}\varrho ^{2}Rd\psi )(-2(d\phi ^{\prime }+\beta )
-\sqrt{2}\varrho ^{2}Rd\psi )-d\psi ^{2}-ds^{2}(M) \ ,
\end{equation}
where $M$ is $\mathbb{S}^{2}$, $\mathbb{R}^{2}$ or $\mathbb{H}^{2}$ and 
\begin{equation}
d\beta =\sqrt{2}\varrho ^{2}\mathrm{dvol}\ (M)\ .
\end{equation}
The scalars $X^{I}$ are constant.

\item[ii)] 
\begin{eqnarray}
ds^{2} &=&u^{4}(dR-\sqrt{2}Rd\psi )^{2}-\sqrt{2}(dR-\sqrt{2}Rd\psi )(Qd\phi
-2u^{-6}dz) \nonumber \\ &-&u^{-2}d\psi ^{2} +({\frac{1}{2}}u^{4}Q^{2}-u^{-2})d\phi ^{2}-{\frac{1}{2}}
u^{-2}(2u^{-6}-Q^{2})^{-1}dQ^{2}
\end{eqnarray}
where the non-constant scalars $X^{I}$, $u$ and $Q$ are constrained by 
\begin{equation*}
u^{-2}X_{I}=\chi QV_{I}+q_{I}
\end{equation*}
for constants $q_{I}$.
\end{itemize}

Our analysis has been entirely local, in particular, we have not assumed
that the horizon is compact. A more detailed investigation of the above solutions,
taking the compactness of the horizon into account, must be undertaken in
order to fully understand these solutions, or to rule them out.
Furthermore, the enhanced local isometries
which these solutions possess appear to depend on the existence of the
additional Killing spinors. It is not clear which, if any, of these extra
isometries exist for black holes preserving only $1/4$ of the supersymmetry.

\appendix

\section{List of Solutions}

In this Appendix, we briefly summarize the formalism of $N=2$, $D=5$
supergravity coupled to vector multiplets, and we also present a 
list of all half supersymmetric solutions,
for which at least one of the Killing spinors generates a timelike Killing
{\normalsize \ }vector. For more details see \cite{halfgs}.

The action of $N=2$, $D=5$ gauged supergravity coupled to $n$
abelian vector multiplets is~\cite{gunaydin:85}
\begin{eqnarray}
S = {\frac{1 }{16 \pi G}} \int \bigg( \star \big(-{}^5 R + 2 \chi^2 {\mathcal{V}} \big)
-Q_{IJ} F^I \wedge \star F^J +Q_{IJ} dX^I \wedge \star dX^J  \nonumber \\
-{\frac{1 }{6}} C_{IJK} F^I \wedge F^J \wedge A^K \bigg)
\end{eqnarray}
where $I,J,K$ take values $1, \ldots ,n$ and $F^I=dA^I$.
The metric has mostly negative signature. $C_{IJK}$ are
constants that are symmetric on $IJK$, $\chi$ is a non-zero constant and 
the gauge field couplings $Q_{IJ}$ are assumed to be invertible, with inverse $Q^{IJ}$.
The $X^I$ are scalars which are constrained via
\begin{equation}  \label{eqn:conda}
{\frac{1 }{6}}C_{IJK} X^I X^J X^K=1\,.
\end{equation}
We may regard the $X^I$ as being functions of $n-1$ unconstrained scalars
$\phi^a$. It is convenient to define
\begin{equation}
X_I \equiv {\frac{1 }{6}}C_{IJK} X^J X^K
\end{equation}
so that the condition~({\ref{eqn:conda}}) becomes
\begin{equation}
X_I X^I =1\,.
\end{equation}
In addition, the coupling $Q_{IJ}$ depends on the scalars via
\begin{equation}
Q_{IJ} = {\frac{9 }{2}} X_I X_J -{\frac{1 }{2}}C_{IJK} X^K
\end{equation}
so in particular
\begin{equation}
Q_{IJ} X^J = {\frac{3 }{2}} X_I\,, \qquad Q_{IJ} d X^J = -{\frac{3
}{2}} d X_I\,.
\end{equation}
The scalar potential can be written as
\begin{equation}
{\mathcal{V}} = 9 V_I V_J (X^I X^J - {\frac{1 }{2}} Q^{IJ})
\end{equation}
where $V_I$ are constants which are not all zero.
There are two sets of Killing spinor equations; the gravitino Killing spinor equation
is
\begin{equation}  \label{eqn:grav}
\left[\nabla_\mu +{\frac{1 }{8}}X_I(\gamma_\mu{}^{\nu \rho} -4 \delta_\mu{}
^\nu \gamma^\rho) F^I{}_{\nu \rho} \right] \epsilon^a-{\frac{\chi }{2}} V_I
(X^I \gamma_\mu-3 A^I{}_\mu) \epsilon^{ab} \epsilon^b=0
\end{equation}
and the dilatino Killing spinor equation is
\begin{eqnarray}  \label{eqn:newdil}
\left[ \left({\frac{1 }{4}}Q_{IJ}-{\frac{3 }{8}}X_I X_J \right) F^J{}_{\mu
\nu} \gamma^{\mu \nu} +{\frac{3 }{4}} \gamma^\mu \nabla_\mu X_I \right]
\epsilon^a 
\nonumber \\
\hskip3cm +{\frac{3 \chi }{2}}\left(X_I V_J X^J -V_I \right) \epsilon^{ab}
\epsilon^b =0\ .
\end{eqnarray}
It is known that for the solutions considered here, the Killing spinor equations,
together with the Bianchi identity $dF^I=0$ are
sufficient to imply that the Einstein, gauge and scalar field equations hold automatically.

All half-supersymmetric solutions of this theory for which at least one of the 
Killing spinor vector bilinears is timelike have been classified in \cite{halfgs}.
The spacetime metric is written as a fibration over a K\"ahler 4-manifold
with metric
\begin{equation}
ds^2 = f^4 \big(dt+\Omega)^2 - f^{-2}ds_{\mathbf{B}}^{2}
\end{equation}
where $ds_{\mathbf{B}}^{2}$ is a K\"{a}hler
4-manifold, $\frac{\partial }{\partial t}$ is a Killing vector which is a
symmetry of the full solution, $f$ is the $t$-independent function which appears in the
definition of the Killing spinors ({\ref{kilsp1}}), and 
$\Omega $ is a $t$-independent 1-form on the K\"{a}hler base $B$.
It was shown in \cite{halfgs} that there are six classes of solution.

\subsection{Solutions of Type (1)}

These solutions fall into two classes, according as to whether a constant of
integration $\theta $ is zero or non-zero. The co-ordinates on the base
space are $\phi ,w,u,v$. If $\theta \neq 0$, then the solution is
{\normalsize \ } 
\begin{eqnarray}
ds_{\mathbf{B}}^{2} &=&H^{2}(d\phi +(v\cos Y+\theta
^{-1}(H^{2}-c^{2}v^{2}f^{-6}))dw)^{2}  \nonumber \\
&+&H^{-2}dv^{2}+H^{2}v^{2}\sin ^{2}Y(dw^{2}+du^{2}),  \nonumber \\
\Omega  &=&-{\frac{1}{2cv}}(H^{2}+c^{2}v^{2}f^{-6})d\phi   \nonumber \\
&-&\big({\frac{1}{2c\theta v}}(H^{4}-c^{4}v^{4}f^{-12})+{\frac{1}{c}}
(H^{2}\cos Y+{\frac{\theta v}{2}})\big)dw,  \nonumber \\
F^{I} &=&d\big(f^{2}X^{I}(dt-{\frac{1}{2cv}}(H^{2}-c^{2}v^{2}f^{-6})d\phi  
\nonumber \\
&-&({\frac{1}{2cv\theta }}(H^{2}-c^{2}v^{2}f^{-6})^{2}+{\frac{H^{2}}{c}}\cos
Y+{\frac{\theta v}{2c}}+cv^{2}f^{-6})dw)\big),  \nonumber \\
X_{I} &=&f^{2}\left( {\frac{q_{I}}{v}+\frac{\chi }{c}}\left( 
\frac{c^{2}v}{f^{6}\theta }-{\frac{H^{2}}{\theta v}}-1\right) V_{I}\right) \ ,
\label{bc1}
\end{eqnarray}
for constant $c \neq 0$ and constants $q_I$, where 
$Y,H$ are functions of 
$u,v$ ($\sin Y\neq 0$) satisfying the constraints 
\begin{eqnarray}
{\frac{\partial H^{2}}{\partial u}} &=&H^{2}v\sin ^{2}Y\left( 3\frac{\chi
cvV_{I}X^{I}}{f^{4}}-\theta \right) ,  \nonumber \\
{\frac{\partial H^{2}}{\partial v}} &=&-\frac{cv}{f^{4}}\left( 3\chi
V_{I}X^{I}+\frac{c}{f^{2}}\right) +\cos Y(3\frac{\chi cvV_{I}X^{I}}{f^{4}}
-\theta ),  \label{Hcons2}
\end{eqnarray}
and{\normalsize \ 
\begin{eqnarray}
{\frac{\partial Y}{\partial u}} &=&\sin Y\left( -H^{2}+3\frac{\chi
cv^{2}V_{I}X^{I}}{f^{4}}+\frac{c^{2}v^{2}}{f^{6}}\right) +\frac{v}{2}\sin
2Y\left( 3\frac{\chi cvV_{I}X^{I}}{f^{4}}-\theta \right) ,  \nonumber \\
{\frac{\partial Y}{\partial v}} &=&-\frac{1}{H^{2}}\sin Y\left( 3\frac{\chi
cvV_{I}X^{I}}{f^{4}}-\theta \right) .  \label{Ycons2}
\end{eqnarray}
}

If $\theta =0$, then the solution is given by{\normalsize \ }
\begin{eqnarray}
ds_{\mathbf{B}}^{2} &=&H^{2}(d\phi +(v(\cos Y-1)-{\frac{c}{\chi }}X)dw)^{2} 
\nonumber \\
&+&H^{-2}dv^{2}+H^{2}v^{2}\sin ^{2}Y(dw^{2}+du^{2}),  \nonumber \\
\Omega  &=&-{\frac{1}{2cv}}(H^{2}+c^{2}v^{2}f^{-6})d\phi   \nonumber \\
&+&\big({\frac{1}{2cv}}(H^{2}+c^{2}v^{2}f^{-6})(v+{\frac{c}{\chi }}X)-
{\frac{H^{2}}{c}}\cos Y\big)dw,  \nonumber \\
F^{I} &=&d\big(f^{2}X^{I}(dt-{\frac{1}{2cv}}(H^{2}-c^{2}v^{2}f^{-6})d\phi  
\nonumber \\
&+&({\frac{1}{2cv}}(H^{2}-c^{2}v^{2}f^{-6})(v+{\frac{c}{\chi }}X)-
{\frac{H^{2}}{c}}\cos Y-cv^{2}f^{-6})dw)\big),  \nonumber \\
vf^{-2}X_{I} &=&XV_{I}+q_{I},
\end{eqnarray}
for constants $c$, $q_{I}$ ($c \neq 0$, $q_{\tilde{I}}=0$, $V_{\tilde{I}} \neq 0$ and $X=v f^{-2} {X_{\tilde{I}} \over V_{\tilde{I}}}$ 
for some fixed ${\tilde{I}}$). 
$Y, H$ are functions of 
$u,v$ ($\sin Y\neq 0$) satisfying the constraints (\ref{Hcons2}) and 
(\ref{Ycons2}) with $\theta =0$.  $X$ is also a function of $u$ and $v$, and satisfies
\begin{eqnarray}
\label{scalcon1}
{\partial X \over \partial v} = {\chi \over c} (\cos Y -1), \qquad {\partial X \over \partial u} = {\chi \over c} H^2 v \sin^2 Y \ .
\end{eqnarray}

\subsection{Solutions of Type (2)}

One can choose a co-ordinate $v$ on $\mathbf{B}$ together with three 
$v$-independent 1-forms $\sigma ^{i}$ ($i=1,2,3$) on $\mathbf{B}$ orthogonal to 
$\frac{\partial }{\partial v}$. There are constants $c$, $\theta $ 
($c\neq 0$) and the solution takes one of three types according as $c\theta $ is
negative, zero or positive. If $\theta \neq 0$ then 
\begin{equation}
ds_{\mathbf{B}}^{2}={\frac{1}{\theta v+c^{2}v^{2}f^{-6}}}dv^{2}
+{\frac{v}{\theta ^{2}}}(\theta +c^{2}vf^{-6})(\sigma ^{1})^{2}+
{\frac{v}{|\theta |}}((\sigma ^{2})^{2}+(\sigma ^{3})^{2}),
\end{equation}
and if $\theta =0$, 
\begin{equation}
ds_{\mathbf{B}}^{2}={\frac{1}{c^{2}v^{2}f^{-6}}}dv^{2}+
4c^{8}f^{-6}v^{2}(\sigma ^{1})^{2}+2c^{3}v((\sigma ^{2})^{2}+(\sigma ^{3})^{2}).
\end{equation}

The 1-forms $\sigma ^{i}$ satisfy 

\begin{tabular}{lr}
$d\sigma ^{i}=-{\frac{1}{2}}\epsilon _{ijk}\sigma ^{j}\wedge \sigma ^{k}$ & 
: \qquad\ \ \textrm{if } $c\theta >0$ \\ 
$d\sigma ^{1}=\sigma ^{2}\wedge \sigma ^{3}$, $d\sigma ^{2}=\sigma
^{1}\wedge \sigma ^{3}$, $d\sigma ^{3}=-\sigma ^{1}\wedge \sigma ^{2}$ & :
\qquad\ \ \textrm{if } $c\theta <0$ \\ 
$d\sigma ^{1}=\sigma ^{2}\wedge \sigma ^{3}$, $d\sigma ^{2}=d\sigma ^{3}=0$
\ \  & : \qquad\ \ \textrm{if } $c\theta =0$ \\ 
& 
\end{tabular}

If $\theta \neq 0$ then 
\begin{equation}
\Omega =-{\frac{cv}{\theta }}f^{-6}\sigma ^{1},
\end{equation}
whereas if $\theta =0$ then 
\begin{equation}
\Omega =2c^{4}vf^{-6}\sigma ^{1}.
\end{equation}

In all cases, the scalars $f$ and $X^{I}$ are constrained by 
\begin{equation}
X_{I}={\frac{f^{2}}{c}}(-2\chi V_{I}+{\frac{\rho _{I}}{\sqrt{2}v}})
\end{equation}
for constants $\rho _{I}$ and 
\begin{equation}
F^{I}=d(f^{2}X^{I}dt).
\end{equation}

\subsection{Solutions of Type (3)}

One can again choose a co-ordinate $v$ on $\mathbf{B}$ together with three $v
$-independent 1-forms $\sigma ^{i}$ ($i=1,2,3$) on $\mathbf{B}$, orthogonal
to $\frac{\partial }{\partial v}$. For these solutions, the scalars $X^{I}$
are constant, and it is convenient to define 
\begin{equation}
\Lambda =c\theta +9\sqrt{2}\chi ^{2}(X^{I}X^{J}-{\frac{1}{2}}
Q^{IJ})V_{I}V_{J}
\end{equation}
for constants $c$, $\theta $ ($c\neq 0$). The scalar $f$ is given by 
\begin{equation}
f^{2}=\sqrt{2}cv.
\end{equation}
The solution takes one of three types according as the constant $\Lambda $
is negative, zero or positive. If $\Lambda \neq 0$ then 
\begin{eqnarray}
ds_{\mathbf{B}}^{2} &=&{\frac{1}{\left( {\frac{1}{2\sqrt{2}cv}}
-\theta v+{\frac{3\chi }{c}}V_{I}X^{I}\right) }}dv^{2}+
{\frac{c^{2}}{\Lambda ^{2}}}\left( {\frac{1}{2\sqrt{2}cv}}-\theta v+{\frac{3\chi }{c}}V_{I}X^{I}\right)
(\sigma ^{1})^{2}  \nonumber \\
&+&{\frac{cv}{|\Lambda |}}\left( (\sigma ^{2})^{2}+(\sigma ^{3})^{2}\right) ,
\end{eqnarray}
and if $\Lambda =0$, 
\begin{eqnarray}
ds_{\mathbf{B}}^{2} &=&{\frac{1}{({\frac{1}{2\sqrt{2}cv}}-\theta v+
{\frac{3\chi }{c}}V_{I}X^{I})}}dv^{2}+2c^{2}({\frac{1}{2\sqrt{2}cv}}-\theta v+
{\frac{3\chi }{c}}V_{I}X^{I})(\sigma ^{1})^{2}  \nonumber \\
&+&\sqrt{2}cv((\sigma ^{2})^{2}+(\sigma ^{3})^{2}),
\end{eqnarray}

The 1-forms $\sigma ^{i}$ satisfy 

{\normalsize 
\begin{tabular}{lr}
$d\sigma ^{i}=-\frac{1}{2}\epsilon _{ijk}\sigma ^{j}\wedge \sigma ^{k}$ & :
\qquad\ \ if  $\Lambda >0$ \\ 
$d\sigma ^{1}=\sigma ^{2}\wedge \sigma ^{3}$, $d\sigma ^{2}=\sigma
^{1}\wedge \sigma ^{3}$, $d\sigma ^{3}=-\sigma ^{1}\wedge \sigma ^{2}$ & :
\qquad\ \ if  $\Lambda <0$ \\ 
$d\sigma ^{1}=\sigma ^{2}\wedge \sigma ^{3}$, $d\sigma ^{2}=d\sigma ^{3}=0$
\ \  & : \qquad\ \ if  $\Lambda =0$ \\ 
& 
\end{tabular}
}

If $\Lambda \neq 0$ then 
\begin{eqnarray}
\Omega  &=&{\frac{1}{\Lambda cv^{2}}}({\frac{1}{2\sqrt{2}}}
+{\frac{3\chi v}{2}}V_{I}X^{I})\sigma ^{1},  \nonumber \\
F^{I} &=&d\left( \sqrt{2}cvX^{I}dt+
{\frac{3\chi }{\sqrt{2}\Lambda }}(Q^{IJ}-X^{I}X^{J})V_{J}\sigma ^{1}\right) ,
\end{eqnarray}
whereas if $\Lambda =0$, then 
\begin{eqnarray}
\Omega  &=&{\frac{\sqrt{2}}{cv^{2}}}({\frac{1}{2\sqrt{2}}}
+{\frac{3\chi v}{2}}V_{I}X^{I})\sigma ^{1},  \nonumber \\
F^{I} &=&d\left( \sqrt{2}cvX^{I}dt+3\chi (Q^{IJ}-X^{I}X^{J})V_{J}\sigma
^{1}\right) .
\end{eqnarray}

\subsection{Solutions of Type (4)}

For the fourth class of solution, the scalars $X^{I}$ are constant 
($V_{I}X^{I}\neq 0$), and{\normalsize \ 
\begin{equation}
f=1.
\end{equation}
} The K\"{a}hler base metric is the product of two 2-manifolds 
\begin{equation}
ds_{\mathbf{B}}^{2}=ds^{2}(M_{1})+ds^{2}(M_{2})
\end{equation}
where $M_{1}$ is $\mathbb{H}^{2}$ with Ricci scalar $R=-18\chi
^{2}(V_{I}X^{I})^{2}$, and $M_{2}$ is $\mathbb{H}^{2}$, $\mathbb{R}^{2}$ or 
$\mathbb{S}^{2}$ with Ricci scalar $R=18\chi ^{2}(Q^{IJ}-X^{I}X^{J})V_{I}V_{J}$.

In addition, we have

\begin{equation}
d\Omega =3\chi V_{I}X^{I}\mathrm{dvol\ }(M_{1}),\qquad F^{I}=3\chi
(X^{I}X^{J}-Q^{IJ})V_{J}\mathrm{dvol\ }(M_{2})
\end{equation}
where $dvol\ (M_{1})$, $dvol\ (M_{2})$ are the volume forms of $M_{1}$, 
$M_{2}$.

\subsection{Solutions of Type (5)}

For the fifth class of solutions, there are two types of solution according
as $(\mathrm{Im}\ \lambda )^{2}+(\mathrm{Im}\ \sigma )^{2}\neq 0$, or 
$\mathrm{Im}\ \lambda =\mathrm{Im}\ \sigma =0$. Then if $(\mathrm{Im}\
\lambda )^{2}+(\mathrm{Im}\ \sigma )^{2}\neq 0$, the co-ordinates on the
base are $\phi ,z,\psi ,Q$ (to be distinguished from the gauge coupling $Q_{IJ}$), and 
\begin{eqnarray}
ds_{\mathbf{B}}^{2} &=&e^{\sqrt{2}\varrho \psi }\bigg[(d\phi 
-{\frac{Q}{\varrho ^{3}}}dz)^{2}+d\psi ^{2}+(2u^{-6}\varrho ^{-2}-\varrho
^{-6}(Q^{2}-\xi ))dz^{2}  \nonumber \\
&+&{\frac{1}{2(2u^{-6}\varrho ^{8}-\varrho ^{4}(Q^{2}-\xi ))}}dQ^{2}\bigg], 
\nonumber \\
u^{-2}X_{I} &=&{\frac{\chi }{\varrho ^{4}}}QV_{I}+q_{I},  \nonumber \\
\Omega  &=&-{\frac{1}{\sqrt{2}\varrho ^{2}}}e^{-\sqrt{2}\varrho ^{2}\psi
}(Qd\phi -{\frac{1}{\varrho ^{3}}}(\xi +2\varrho ^{2}u^{-6})dz),  \nonumber \\
F^{I} &=&d\left( u^{2}e^{\sqrt{2}\varrho ^{2}\psi }X^{I}(dt+\Omega )\right) \nonumber \\
&+&3\sqrt{2}\chi \varrho ^{-5}u^{-2}V_{I}(X^{I}X^{J}-
{\frac{1}{2}}Q^{IJ})dz\wedge dQ,  \nonumber \\
f &=&e^{{\frac{\varrho ^{2}}{\sqrt{2}}}\psi }u,
\end{eqnarray}
for constants $q_{I}$, $\xi $, $\varrho \neq 0$. It should be noted that the function
$u$ is defined as a function of $Q$ by the second condition in this expression, on using the
condition $X_I X^I=1$ obtained from the Very Special geometry of the scalar manifold.

It is also convenient to define 
\begin{equation}
\mathcal{G}={\frac{2i}{u}}e^{{\frac{\varrho ^{2}}{\sqrt{2}}}\psi }
\mathrm{Im\ }\lambda ,\qquad \mathcal{H}={\frac{2i}{u}}
e^{{\frac{\varrho ^{2}}{\sqrt{2}}}\psi }\mathrm{Im\ }\sigma ,
\end{equation}
then $\mathcal{H}$ and $\mathcal{G}$ are related to $Q$ by 
\begin{equation}
Q^{2}=\xi +\varrho ^{6}({\frac{2u^{-6}}{\varrho ^{2}}}+\mathcal{G}^{2}+
\mathcal{H}^{2}).  \label{q2con1}
\end{equation}
If, however, $\mathrm{Im}\ \lambda =\mathrm{Im}\ \sigma =0$, then the
solutions have constant $X^{I}$, and 
\begin{eqnarray}
ds_{\mathbf{B}}^{2} &=&e^{\sqrt{2}\varrho ^{2}\psi }((d\phi +\beta
)^{2}+d\psi ^{2}+ds^{2}(M)),\qquad d\beta =\sqrt{2}\varrho ^{2}\mathrm{dvol}
\ (M)  \nonumber \\
{}^{(M)}R &=&4\varrho ^{4}-36\chi ^{2}(X^{I}X^{J}-{\frac{1}{2}}
Q^{IJ})V_{I}V_{J},  \nonumber \\
\Omega  &=&-3\chi V_{I}X^{I}
{\frac{e^{-\sqrt{2}\varrho ^{2}\psi }}{\sqrt{2}\varrho ^{2}}}(d\phi +\beta ),  \nonumber \\
F^{I} &=&d\left( e^{\sqrt{2}\varrho ^{2}\psi }X^{I}(dt+\Omega )\right)
+6\chi (X^{I}X^{J}-{\frac{1}{2}}Q^{IJ})V_{J}\mathrm{dvol}\ (M),  \nonumber \\
f &=&e^{{\frac{\varrho ^{2}}{\sqrt{2}}}\psi },
\end{eqnarray}
where $\varrho $ is a non-zero constant. $M$ is a 2-manifold which is either 
$\mathbb{S}^{2}$, $\mathbb{R}^{2}$ or $\mathbb{H}^{2}$.

\subsection{Solutions of Type (6)}

For the sixth class of solutions, there are two types of solution according
as $(\mathrm{Im}\ \lambda )^{2}+(\mathrm{Im}\ \sigma )^{2}\neq 0$, or 
$\mathrm{Im}\ \lambda =\mathrm{Im}\ \sigma =0$. Then if $(\mathrm{Im}\
\lambda )^{2}+(\mathrm{Im}\ \sigma )^{2}\neq 0$, the co-ordinates on the
base are $\phi ,\psi, y ,z$ and 
\begin{eqnarray}
f^{-2}X_{I} &=&XV_{I}+q_{I},\quad \mathrm{where}\ X\ \mathrm{satisfies}\quad 
{\frac{1}{4\chi }}{\frac{dX}{dy}}-{\frac{1}{2}}f^{-6}=
{\frac{\xi }{\sqrt{2}\chi }}X,  \nonumber \\
ds_{\mathbf{B}}^{2} &=&d\phi ^{2}+d\psi ^{2}+2\sqrt{2}\left( {\frac{1}{2}}
f^{-6}+{\frac{\xi }{\sqrt{2}\chi }}X\right) (dy^{2}+dz^{2}),  \nonumber \\
d\Omega  &=&\xi d\phi \wedge d\psi +d\left( (\sqrt{2}f^{-6}+
{\frac{\xi }{\sqrt{2}\chi }}X)dz\right) ,  \nonumber \\
F^{I} &=&d\bigg(f^{2}X^{I}(dt+\Omega )-\sqrt{2}f^{-4}X^{I}dz\bigg) \ ,
\end{eqnarray}
for constants $\xi, q_I$. In this solution, the vector field $K$ is given by
\begin{equation}
K = {\partial \over \partial \phi}
\end{equation}
and hence $K^2=1$.

If, however, $\mathrm{Im}\ \lambda =\mathrm{Im}\ \sigma =0$, the scalars 
$X^{I}$ are constant, and
\begin{equation}
f=1.
\end{equation}
Then 
\begin{equation}
ds_{\mathbf{B}}^{2}=d\phi ^{2}+d\psi ^{2}+ds^{2}(M),
\end{equation}
where $M$ is a 2-manifold which is either $\mathbb{S}^{2}$, $\mathbb{R}^{2}$
or $\mathbb{H}^{2}$ according as the Ricci scalar 
\begin{equation}
{}^{(M)}R=-36\chi ^{2}(X^{I}X^{J}-{\frac{1}{2}}Q^{IJ})V_{I}V_{J}
\end{equation}
is positive, zero, or negative. In addition, one has 
\begin{eqnarray}
d\Omega  &=&-3\chi V_{I}X^{I}(\mathrm{dvol\ }(M)+d\phi \wedge d\psi ), 
\nonumber \\
F^{I} &=&-3\chi X^{I}X^{J}V_{J}d\phi \wedge d\psi +3\chi
(X^{I}X^{J}-Q^{IJ})V_{J}\mathrm{dvol\ }(M).
\end{eqnarray}
Again, for this solution 
\begin{equation}
K = {\partial \over \partial \phi}
\end{equation}
and hence $K^2=1$.

\section*{Acknowledgments}

The work of W. A. Sabra was supported in part by the National Science
Foundation under grant number PHY-0703017. J. Gutowski thanks H. Kunduri and
H. Reall for many useful discussions.

\end{document}